\documentclass[aps,prl,floatfix,tightenlines,superscriptaddress,twocolumn]{revtex4}
\usepackage{amssymb, amsfonts, latexsym, graphicx, natbib, bbm,
multirow, textcomp, ulem, floatflt, amsmath}
\usepackage{color}
\usepackage[bookmarks = true, colorlinks = true, urlcolor=blue]{hyperref}
\def\>{\rangle}
\def\<{\langle}
\def\rmeas{\ensuremath{\rho_{{S_1}{S_2}}}}

\newcommand{\etal}{\textit{et.~al.}}
\newcommand{\hsp}[1]{\hspace{#1 em}}
\newcommand{\sqz}{\hsp{-0.1}}
\def\iden{\mathbbmss{1}}
\newcommand{\ketbra}[2]{\left\vert{#1}\right\rangle \sqz\sqz\sqz \left\langle{#2}\right\vert}

\newcommand{\ket}[1]{\left\vert{#1}\right\rangle}

\newcommand{\dg}{\ensuremath{^{\circ}}}
\newcommand{\Exp}[1]{\ensuremath{e^{#1}}}
\graphicspath{{ClusterPOVM_figures/}{Cluster_POVM_Paper/}}

\begin{document}

\title{Cluster-state quantum computing enhanced by high-fidelity generalized measurements}
\author{D.N. Biggerstaff}
\affiliation{Institute for Quantum Computing and Department of Physics \&
Astronomy, University of Waterloo, Waterloo, Canada, N2L 3G1}
\author{T. Rudolph}
\affiliation{QOLS, Blackett Laboratory, Imperial College London, Prince Consort Road, London SW7 2BW, United Kingdom}
\author{R. Kaltenbaek}
\affiliation{Institute for Quantum Computing and Department of Physics \&
Astronomy, University of Waterloo, Waterloo, Canada, N2L 3G1}
\author{D. Hamel}
\affiliation{Institute for Quantum Computing and Department of Physics \&
Astronomy, University of Waterloo, Waterloo, Canada, N2L 3G1}
\author{G. Weihs}
\affiliation{Institut f\"ur Experimentalphysik, Universit\"at Innsbruck, Technikerstr. 25, 6020 Innsbruck, Austria}
\affiliation{Institute for Quantum Computing and Department of Physics \&
Astronomy, University of Waterloo, Waterloo, Canada, N2L 3G1}
\author{K.J. Resch}
\affiliation{Institute for Quantum Computing and Department of Physics \&
Astronomy, University of Waterloo, Waterloo, Canada, N2L 3G1}

\begin{abstract}
\noindent We introduce and implement a technique to
extend the quantum computational power of cluster states
by replacing some projective measurements with generalized
quantum measurements (POVMs).
As an experimental demonstration we fully realize an arbitrary three-qubit cluster computation by implementing a tunable linear-optical POVM, as well
as fast active feedforward, on a two-qubit photonic cluster state.
Over 206 different computations, the average output fidelity is $0.9832\pm0.0002$; furthermore the error contribution from our POVM device and feedforward 
is only of $\mathit{O}(10^{-3})$, less than some recent thresholds for fault-tolerant cluster computing.
\end{abstract}

\maketitle

Measurement-based (cluster) computation 
\cite{RaussendorfBriegelBrowne}
is an attractive alternative to standard circuit-based quantum
computing. Instead of requiring multi-qubit gates, which are
hard to implement experimentally, cluster computing
requires only simple, single-qubit projective measurements.
However, the prerequisite is a highly-entangled, multi-qubit
cluster state. Thus far, laboratory cluster states \cite{Walther2005,Kiesel2005,Prevedel2007,Lu2007,Chen2007,VallonePapers}
have proven difficult to generate and limited in size.
In order to make the most of these resources it is thus
desirable to find means to extend
the computational power of available clusters.
Here we introduce 
 one such technique, based on
performing Positive Operator-Valued Measures (POVMs)\cite{Peres1993, POVM} 
on cluster-state qubits instead of standard projective measurements.
As an experimental demonstration, we implement this technique 
to perform a three-qubit cluster
computation for state preparation using linear optics and two entangled
photons. 
Our results show 
the error introduced by the POVM apparatus
and subsequent feedforward 
to be of $O( 10^{-3})$, suggesting operation within recent thresholds
for fault-tolerant cluster quantum computing \cite{Raussendorf2007}.
\begin{figure}
\includegraphics[width=1 \columnwidth]{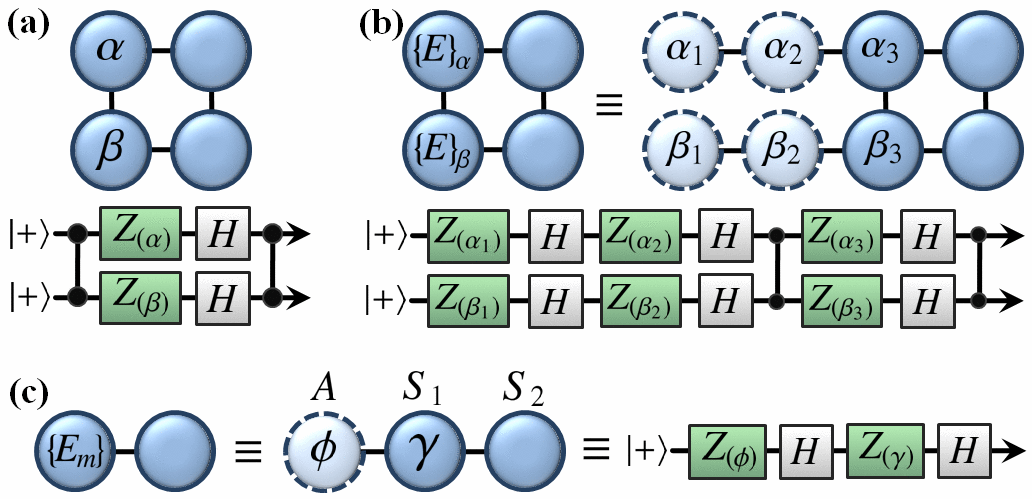}
\caption{(Color online): Cluster computing with POVMs. (a) A
four-qubit `box' cluster, one of the largest clusters achievable experimentally
with optical qubits 
\cite{Walther2005,Kiesel2005,Prevedel2007,Lu2007,Chen2007,VallonePapers}. 
By measuring the input cluster qubits and implementing the correct
Pauli error correction, one achieves output equivalent to 
 the circuit
shown. Note: $\{X,Y,Z\}$: standard Pauli operators; $H$: Hadamard operator $(X+Z)/\sqrt{2}$; $Z_{(\beta)}$: $\exp{\left(iZ\beta/2\right)}$; $\alpha$ (on a cluster qubit): measurement in $\ket{\alpha_\pm} = \ket{0}\pm \exp{(i\alpha)}\ket{1}$. (b) A larger computation, and equivalent circuit, achievable via implementing POVMs on
each input qubit of the same `box' cluster. Cluster qubits with dashed outlines
represent `virtual' qubits simulated by performing POVMs $\{E\}_\alpha$ and
$\{E\}_\beta$.  (c) The computation we perform experimentally. By performing
the POVM $\{E_m\}$ on the input qubit to 
our two-qubit cluster, we 
 realize a three-qubit linear cluster computation, which in turn 
corresponds to a circuit for arbitrary pure state preparation via successive rotations of $\ket{+}$
around the $z$ and $x$ axes.
\label{fig:Theory}
}
\end{figure}

Every POVM
can be implemented by interacting an ancilla 
with the system to be measured,
and performing projective measurement(s) on the
combined Hilbert space \cite{Peres1993}.  Consider
using the controlled-$Z$ (CZ) operation (the interaction) to attach a new small cluster (the ancilla) to a qubit $S$ (the system) in a cluster to be used for computation. By then performing single-qubit projective measurements on the ancillary cluster and $S$, we implement a POVM on $S$. 
For cluster computation, it matters only
that the correct POVM is implemented; the manner in which this is accomplished
is immaterial. In particular it may not be necessary to perform
the (often technically challenging) task of interacting 
$S$ with
ancillary physical systems; the same POVM can be implemented
using additional 
degrees of freedom of the qubit itself
\cite{PreskillsNotes,Lanyon}, which are readily available in many 
architectures.
For the photonic polarization qubits considered herein, one can employ
an additional spatial-mode degree of freedom to implement an arbitrary 4-outcome POVM, and all single-qubit POVMs can be obtained by suitable (classical) processing of such a POVM \cite{PreskillsNotes}.

As a specific example, consider the 4-qubit box cluster \cite{Walther2005}, and equivalent circuit, shown in Fig. \ref{fig:Theory}a).  By performing POVMs on each of the first two 
qubits, one can effectively add `virtual' qubits, thereby simulating a larger circuit, as shown in Fig. \ref{fig:Theory}b).

Our approach differs fundamentally from recent
experiments employing hyperentangled photon pairs for cluster 
computing \cite{VallonePapers,Chen2007}.  In contrast to those
works, our method can replace arbitrarily large pieces of
cluster, and avoids the complications of sources producing particles entangled in multiple degrees of freedom.
Furthermore our technique incorporates perfect automatic feedforward in the `virtual' qubits, and is sufficiently versatile to serve as a useful primitive for large-scale cluster computers: POVMs (including those performed herein) can be applied, without modification, towards enhancing the computational power of \textit{any} given cluster.

Before 
addressing the specifics of implementing an optical POVM
to simplify cluster computations, let us mention some other
closely related
quantum information processing tasks. 
The ability to perform an arbitrary POVM on one qubit from a (not
necessarily maximally) entangled pair constitutes the basic
primitive of quantum steering \cite{Schrodinger}, which 
underlies the optimal cheating attacks \cite{BB84} on
generalizations of the BB84 two-party bit commitment protocol.
It also underlies the procedure for achieving maximum
disturbance-free control \cite{Spek2}, a basic primitive of quantum 
cryptography.

As a demonstration of our technique, consider the computation in
Fig.~\ref{fig:Theory}c). We employ the smallest non-trivial cluster of two qubits; as depicted, a POVM on the first 
qubit allows a three-qubit computation. 
We label the
ancillary qubit $A$ and the system qubits $S_1$ and $S_2$. In the manner of a standard cluster computation, we imagine $A$ is initially in the state
$|+\rangle_A$, is bonded to $S_1$ via a $CZ$ gate $CZ_{S_1 A}$, and
is then projected into the basis
$\ket{\phi_{\pm}}=\ket{0}\pm e^{i\phi}\ket{1}$.
Depending on the sign of the outcome ($+$ or $-$), qubit $S_1$ must subsequently be projected into either
$\ket{\gamma_\pm}$ or 
$X\ket{\gamma_\pm}$, respectively.
 These measurements are equivalent to the 4 projectors
$\Pi_{ab}^{{S_1} A}=Z^a|\phi_+\>_A\<\phi_+|Z^a \otimes
\linebreak[0]
X^aZ^b|\gamma_+\>_{S_1}\<\gamma_+|Z^bX^a$
, where $a,b\in\{0,1\}$.
We can represent this process as a POVM on $S_1$
as:
\begin{equation}
\label{eq:Projectors}
E_{ab}^{S_1} =\<+_A|CZ_{{S_1} A}\Pi_{ab}^{{S_1} A}CZ_{{S_1} A}|+_A\>.
\end{equation}
As can be readily verified, the POVM elements are 
$\{E_m\} = \frac{1}{2}\left\{\sigma_m \ketbra{\chi}{\chi} \sigma_m^{\dagger} \right\}$, 
where we use the index $m=1,..4$, $\ket{\chi(\phi,\gamma)}=\cos{(\phi/2)}\ket{0}-i \Exp{i \gamma}\sin{(\phi/2)} \ket{1}$, and $\sigma_m \in \{\iden ,X,XZ,Z\}$. 

The cluster model requires active feedforward in order to drive
deterministic quantum computations despite inherently random
measurement outcomes \cite{Pittman,Prevedel2007}. When using POVMs on cluster qubits, the required feedforward depends on the measurement outcome: after performing POVM $\{E_m\}$ on $S_1$ and obtaining outcome $m$, the operator $\sigma_m$ must be applied to the output qubit $S_2$ in order to recover the outcome of the circuit in Fig. \ref{fig:Theory}c).
	
In our experiment we begin with photons in the Bell state 
$\ket{\Phi^+}_{{S_1}{S_2}}$, where $\ket{\Phi^{\pm}}=\ket{HH}\pm\ket{VV}$,
and $H$ and $V$ indicate horizontal and vertical polarization, our
 $\ket{0}$ and $\ket{1}$, respectively. 
$\ket{\Phi^+}$ differs from a 2-qubit cluster by a Hadamard on one photon. We make a convenient adjustment of our second measurement angle so as to implement the POVM:
\begin{equation} \label{eq:POVM}
\{\mathcal{E}_n(\phi,\,\theta)\} =
\frac{1}{2}\{Z \rho^* Z,\,
\rho^* ,\,
X \rho^* X ,\,
XZ \rho^* ZX
\},
\end{equation}
where $\rho=\ketbra{\psi}{\psi}$, and
$\ket{ \psi 	( \phi , \,\theta )} =
	\cos{	(	\phi/2 ) } \ket{H} +
	\Exp{i \theta}  \sin{ ( \phi/2 ) } \ket{V}$. After performing $\{\mathcal{E}_n\}$ on photon $S_1$ from our actual 2-photon state $\ket{\Phi^+}_{{S_1}{S_2}}$, obtaining outcome $n$, and implementing $\sigma_n \in \{Z,\iden,X,XZ\}$ on $S_2$, the output will be state $\ket{\psi(\phi,\theta)}$, where $\phi$ and $\theta$ are adjustable experimental parameters which respectively correspond to the polar and azimuthal angle of the output in the Bloch sphere representation.

\begin{figure}
\includegraphics[width=1 \columnwidth]{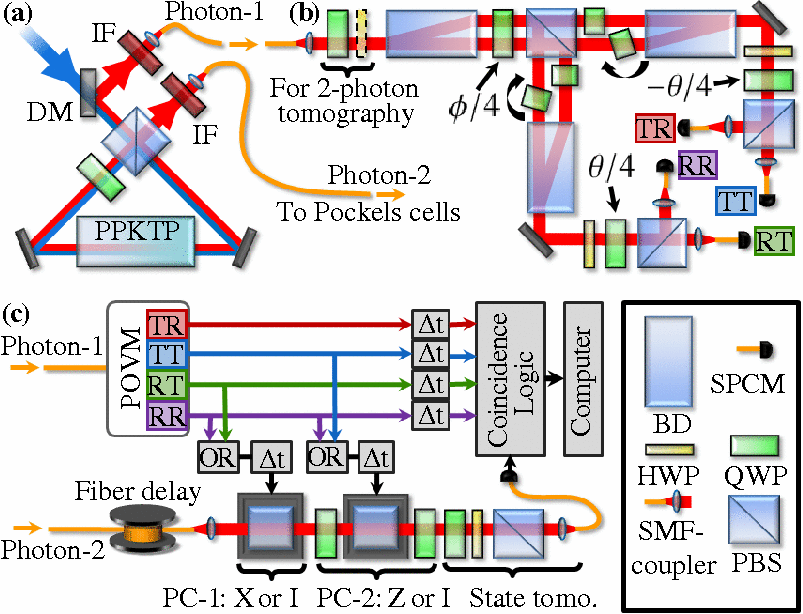}
\caption{
\label{fig:Setup}
(Color online). Experimental implementation of cluster-state computing with a POVM.
a) A source \cite{Kim2006}
produces maximally-entangled photon pairs coupled into
single-mode fiber (SMF).  b) The POVM is based on an optical interferometer constructed using calcite beam-displacers (BDs), which
couple the polarization of the photon to the path, thereby
enlarging the state space for the generalized measurement. 
Details are given in the text.  c) Schematic of the cluster-state quantum
computer. Photon-1 is measured in the POVM. Two Pockels cells (PCs), each fired dependent on the measurement outcome, actively perform the required correction of Pauli errors. A 50~m SMF serves to delay Photon-2, allowing time to trigger the PCs.  The computational output is analyzed using quantum state tomography. Note: DM: dichroic mirror; IF: blocking and interference filter; PBS: polarizing beamsplitter; H(Q)WP: half (quarter) wave-plate; SPCM: fiber-coupled single-photon counting module.
}
\end{figure}

We generate entangled photons 
as shown in Fig.~\ref{fig:Setup}a) 
\cite{Kim2006}. A grating-stabilized diode laser outputs 0.86~mW
at 404.5~nm to bi-directionally pump a 25~mm
periodically-poled KTiOPO$_4$ (PPKTP) crystal
in a polarization-dependent Sagnac interferometer,
yielding 809~nm entangled photons via type-II parametric downconversion.  At this power, the SMFs typically output singles (coincidence) rates of 150 (30)~kHz.
Polarization controllers (bat-ears) in the SMFs
ensure the output is $\ket{\Phi^+}$.

The apparatus for performing the POVM
$\{\mathcal{E}_n\}$ is depicted schematically in Fig.~\ref{fig:Setup}b).  It is a
polarization-based double interferometer employing calcite beam
displacers (BDs). Due to transverse walk-off, these couple 
polarization
with optical path, 
enlarging the state space from dimension two to four. 
The settings $\{\phi,\theta\}$ of the POVM are
determined by 
half-wave plates (HWPs) in the
interferometer.  Polarization measurements are implemented
using polarizing beamsplitters (PBSs).  The four output modes
are coupled into SMF and 
detected using single-photon counting modules (SPCs).  The BD
construction is inherently phase-stable because the
interfering paths propagate through common optics~\cite{Obrien2003}.
Furthermore, we align the setup using an 809~nm diode laser injected
through the input SMF and a removable polarizing optic, and typically
measure classical interference visibilities $>99.8\%$;
thus the setup is promising for high-fidelity, stable
operation. 

The action of this apparatus can be understood as follows. An arbitrary
input 
qubit is in state $a \ket{H}
+ b \ket{V}$. 
The first BD displaces the $H$- relative to the $V$-component, introducing a `path qubit' with basis
states upper $\ket{U}$ and lower $\ket{L}$, and thereby creating the
entangled state $a\ket{HU}+b\ket{VL}$.  The polarization qubit
is then measured using a HWP at angle $\phi/4$ and a PBS.
HWPs at $45\dg$ then flip the
polarization in path $U$ ($L$) in the transmitted (reflected)
arm. 
HWPs at $0\dg$ are included in the other path to balance path-lengths, and allow the phase to be adjusted via tilting 
about their vertical axes. Recombining the paths at 
subsequent BDs converts the `path' qubit back to
polarization;  this qubit is then
measured via a QWP at $45\dg$, a HWP at $-\theta/4$
($+\theta/4$), and a PBS in the
transmitted (reflected) arm. This yields the POVM elements
$\{\mathcal{E}_n\}$ as follows: The outcome TR which stems from
(T)ransmission at the first PBS and (R)eflection at the second
then corresponds to $\rho^* (\phi,\,\theta)/2$,
and the TT, RT and RR outcomes to $Z \rho^* Z/2$, $X \rho^* X/2$, and $XZ\rho^* ZX/2$, respectively.

We implement the necessary feedforward using two fast RbTiOPO$_4$ (RTP) Pockels cells
(PCs) (Leysop 
 RTP4-20-AR800), able to switch to their half-wave voltage of 1.027~kV in $<5$~ns, both oriented such as to
enact an $X$ operation (i.e. a HWP at $45\dg$) when triggered.
The first PC will be triggered by POVM outcomes RT or RR;
the second is 
surrounded by HWPs 
which rotate its action to $Z$, and
is triggered by outcomes TT or RR.  Photon-2 is stored
in a 50m SMF for 250~ns of delay to allow
ample time for detection, logic, and triggering the 
PCs. Note 
that one stage of feedforward is
incorporated directly into the design of the POVM since the 
angle of the last HWP is dependent on the outcome of the 
polarization measurement. This 
reduces experimental cost and complexity, as we require one less
PC over a direct implementation \cite{Prevedel2007}.  It
also improves 
computational speed, as each additional PC requires delaying the relevant photon by $O$(100)ns to allow time for detection, logic, and
triggering; in our POVM this feedforward
requires a mere couple ns of 
optical path. After correction, we perform state tomography on 
Photon-2 using a HWP, QWP, PBS, and 
SPCM. Our raw
data consists of coincidence counts between this output and any
one of the four POVM outcomes.

\begin{figure}
\hspace{-0.03 \columnwidth}
\includegraphics[width=0.47 \columnwidth
	]{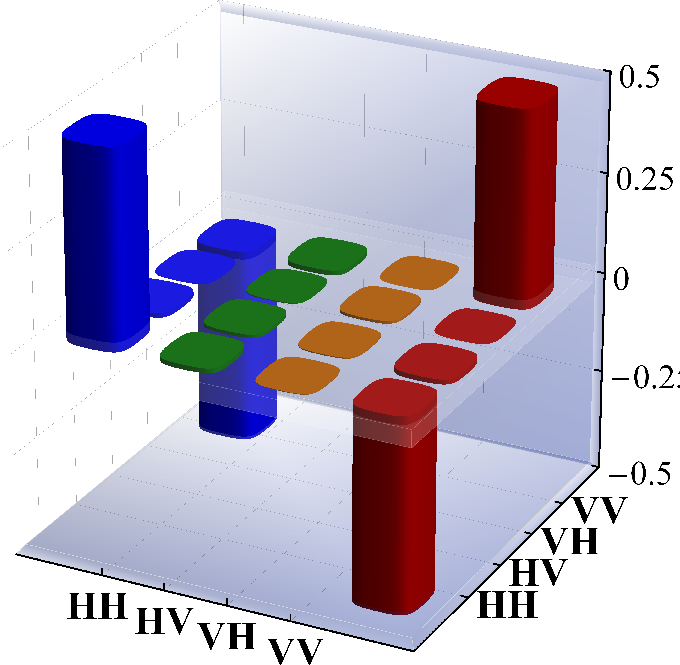}
	\hspace{-0.01 \columnwidth}
\includegraphics[width=0.47 \columnwidth
	]{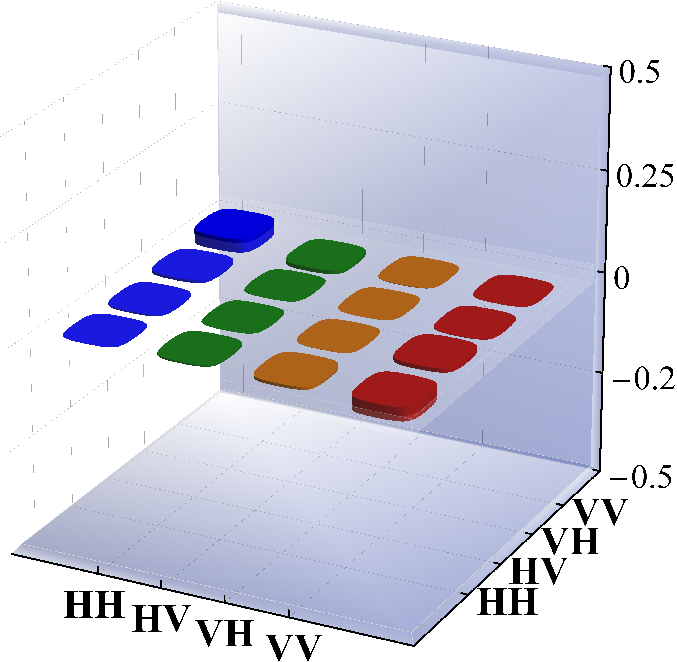}
	\hspace{-0.03 \columnwidth}
\caption{
\label{fig:TwoPhotonDM}
(Color online:) Experimentally reconstructed density matrix \rmeas of 
our 2-photon cluster state: real part (left)
and imaginary part (right).
The fidelity with the ideal state $|\Phi^{-}\rangle$
is $F=0.980\pm0.001$; the tangle is $T=0.926\pm0.002$ and the purity, $\mathrm{Tr}(\rho^2)$, is $P=0.963\pm0.002$ \cite{Jozsa, White2001}.
}
\end{figure}

We characterized the entangled state generated in fibers 1
and 2 (see Fig~\ref{fig:Setup}a) via over-complete 
state tomography. To analyze Photon-1, we employed the first two
waveplates in the POVM and the $H$ 
output of the
first BD; the $V$ output was blocked, and the other waveplates
set so as to direct all photons to output TT.
We then counted coincidences between this output and that of
the polarization analyzer following the (switched off) PCs.
The density matrix is reconstructed via a 
maximum-likelihood technique \cite{James2001}, 
and shown in Fig.~\ref{fig:TwoPhotonDM}.
Our measured state $\rmeas$ has fidelity $F=0.980$ with 
$\ket{\Phi^-}$. When not in use
for tomography, the QWP at the beginning of the POVM is removed
and the HWP set to $0\dg$, which 
maps the $\ket{\Phi^-}$ source state to $\ket{
\Phi^+ }$.

We tested this cluster computer by performing
computations with 206 different measurement settings $\{
\theta,\phi \}$ over a period of 4 hours. 
The target output states $\{\ket{\psi(\phi,\theta)}\}$ include the six
eigenstates of $X$, $Y$ and $Z$, and 
200 numerically-generated
settings designed to be spread evenly over the
surface of the Bloch sphere. For each computation, the output
density matrix $\rho_m$ is tomographically reconstructed 
based on coincidence measurements
integrated over 8~s for each of six analyzer settings
(the eigenstates of $X$, $Y$ and $Z$.) We record
maximum coincidence rates of about 3~kHz, summed
over the four outputs.

\begin{figure}
	\centering
	\includegraphics[width=1 \columnwidth]{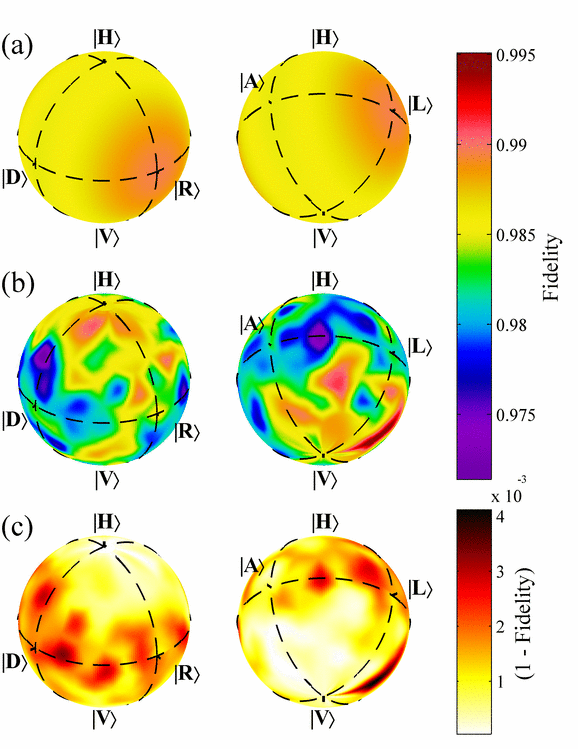}
\caption{
	\label{fig:Results}(Color) Expected and measured fidelity of computational output.
	a) The expected output fidelity based on the measured cluster state, $\rho_{{S_1}{S_2}}$,
assuming perfect POVM operation and feedforward.  b) The measured output fidelity with the target outputs;
the mean is $F = 0.9832\pm0.0002$.
	c) The deviation from unity of the fidelity between the expected and measured
	states $1 - F\left( \rho_e,\,\rho_m \right)$; the mean is
	$(1-F) = (1.16 \pm 0.05) \times 10^{-3}$.  This shows our POVM and feedforward to be operating with very
	high fidelity.
}
\end{figure}

Using the reconstructed 
density matrix $\rmeas$ (Fig.~\ref{fig:TwoPhotonDM}), and assuming a perfect POVM 
and feedforward, we can estimate the 
output 
of the computation: 
$\rho_e\left(\phi,\,\theta\right)= \sum_{n=1}^{4}
\sigma_n \mbox{Tr}_1\sqz\sqz \left[(\mathcal{E}_n \sqz \otimes \sqz \iden) \rmeas (\mathcal{E}_n \sqz \otimes \sqz \iden)^{\dagger}\right]\sigma_n^{\dagger}$.
Fig.~\ref{fig:Results}a) shows the fidelity between $\rho_e$ and the target output $\ket{\psi}$.

Fig.~\ref{fig:Results}b) shows the fidelity
$F\left(\rho_m,\ket{\psi}\right)$ between our measured states
 and the target outputs.
For these 206 computations the mean fidelity is
$0.9832\pm0.0002$, where the uncertainty is the
standard error in the mean. This compares favorably with the expected
mean fidelity $F(\rho_e,\ket{\psi})=0.9865\pm0.0001$ for the same 206 states. The distribution of $F(\rho_m,\ket{\psi})$ is largely explained by two factors: expected variance due to the imperfect entangled state, and fluctuations from
Poissonian counting statistics, as determined by a Monte Carlo 
simulation.

We characterize the errors introduced by the POVM and
feedforward using the quantity
$\left(1-F\left(\rho_e,\,\rho_m\right)\right)$, shown in Fig.~\ref{fig:Results}c), where $F$ is the mixed-state
fidelity \cite{Jozsa}.
With a mean error of only $(1-F) = (1.16 \pm 0.05) \times
10^{-3}$, our results demonstrate remarkable agreement with the
model (Fig.~\ref{fig:Results}a) based on $\rho_{S_1 S_2}$. More
importantly, this shows that our POVM and feedforward are very stable and exhibit a low error rate, comparable to some thresholds for fault-tolerant cluster computing \cite{Raussendorf2007}.

We have shown how POVMs may be employed in cluster
quantum computing to increase the computational power of any given cluster. Furthermore we have experimentally demonstrated this technique by fully realizing an arbitrary, high-fidelity three-qubit cluster computation using two photons. Some
feedforward steps can be incorporated into the design of the POVM, significantly improving computational speed while reducing experimental complexity. This technique
should be incorporated into future cluster computers
to maximize the utility of available resources.

\begin{acknowledgments}
We thank N. Killoran, N. L\"utkenhaus, and K.M. Schreiter  for
valuable discussions, and Z.-W. Wang for designing and building
our TTL logic. D.B. acknowledges financial support from the
Mike and Ophelia Lazaridis Fellowship.  T.R. acknowledges support from EPSRC
and the US Army Research Office. We are grateful for
financial support from NSERC, OCE, and CFI.
\end{acknowledgments}


\begin{thebibliography}
\expandafter\ifx\csname
natexlab\endcsname\relax\def\natexlab#1{#1}\fi
\expandafter\ifx\csname bibnamefont\endcsname\relax
  \def\bibnamefont#1{#1}\fi
\expandafter\ifx\csname bibfnamefont\endcsname\relax
  \def\bibfnamefont#1{#1}\fi
\expandafter\ifx\csname citenamefont\endcsname\relax
  \def\citenamefont#1{#1}\fi
\expandafter\ifx\csname url\endcsname\relax
  \def\url#1{\texttt{#1}}\fi
\expandafter\ifx\csname
urlprefix\endcsname\relax\def\urlprefix{URL }\fi
\providecommand{\bibinfo}[2]{#2}
\providecommand{\eprint}[2][]{\url{#2}}

\bibitem{RaussendorfBriegelBrowne} H.J. Briegel and R. Raussendorf,
	Phys. Rev. Lett. \textbf{86}, 910 (2001); R. Raussendorf and H.J.
	 Briegel, \textit{ibid}. \textbf{86}, 5188 (2001); R. Raussendorf, 
	 D.E. Browne, and H.J. Briegel, Phys. Rev. A. \textbf{68}, 022312
	 (2003).

\bibitem[{\citenamefont{Walther
    et~al.}(2005)\citenamefont{Walther, Resch,
  Rudolph, Schenck, Weinfurter, Vedral, Aspelmeyer, and
  Zeilinger}}]{Walther2005}
P. Walther \etal,  \bibinfo{journal}{Nature} \textbf{\bibinfo{volume}{434}},
  \bibinfo{pages}{169} (\bibinfo{year}{2005}).

\bibitem[{\citenamefont{Kiesel
    et~al.}(2005)\citenamefont{Kiesel, Schmid,
  	Weber, T\'oth, G\"uhne, Ursin, and Weinfurter}}]{Kiesel2005}
	N. Kiesel \etal,
  	\bibinfo{journal}{Phys. Rev. Lett.} \textbf{\bibinfo{volume}{95}},
  	\bibinfo{pages}{210502} (\bibinfo{year}{2005}).

\bibitem[{\citenamefont{Prevedel
    et~al.}(2007)\citenamefont{Prevedel, Walther,
  	Tiefenbacher, Bohi, Kaltenbaek, Jennewein, and Zeilinger}}]{Prevedel2007}
	R. Prevedel \etal,
  	\bibinfo{journal}{Nature} \textbf{\bibinfo{volume}{445}}, \bibinfo{pages}{65}
  	(\bibinfo{year}{2007}).

\bibitem[{\citenamefont{Lu et~al.}(2007)\citenamefont{Lu, Zhou,
    Guhne, bo~Gao,
  	Zhang, sheng Yuan, Goebel, Yang, and Pan}}]{Lu2007}
	C.-Y Lu \etal,
  	\bibinfo{journal}{Nature Physics} \textbf{\bibinfo{volume}{3}},
  	\bibinfo{pages}{91} (\bibinfo{year}{2007}).

\bibitem{Chen2007}K. Chen et.~al., Phys. Rev. Lett. \textbf{99}, 120503 (2007).

\bibitem[{\citenamefont{Vallone
    et~al.}(2007)\citenamefont{Vallone, Pomarico,
 	 Mataloni, Martini, and Berardi}}]{VallonePapers}
	G. Vallone \etal,
  	\bibinfo{journal}{Phys. Rev. Lett.} \textbf{\bibinfo{volume}{98}},
  	\bibinfo{eid}{180502} (\bibinfo{year}{2007});
G. Vallone \etal,
  \bibinfo{journal}{Phys. Rev. A}
  \textbf{\bibinfo{volume}{78}}, \bibinfo{eid}{042335}
  (\bibinfo{year}{2008}).

\bibitem[{\citenamefont{Peres}(1993)}]{Peres1993}
    \bibinfo{author}{\bibfnamefont{A.}~\bibnamefont{Peres}},
  \emph{\bibinfo{title}{Quantum theory: concepts and methods}}
  (\bibinfo{publisher}{Dordrecht: Kluwer}, \bibinfo{year}{1993}), ISBN
  \bibinfo{isbn}{0792325494}.
  
\bibitem{POVM}S.M. Barnett and E. Riis, J. Mod. Opt. \textbf{44}, 1061 (1997); 
	S.E. Ahnert and M.C. Payne, Phys. Rev. A. \textbf{71}, 012330 (2005).

\bibitem{Raussendorf2007}
		R. Raussendorf, J. Harrington and K. Goyal, New J. Phys. \textbf{9},
		199 (2007).

\bibitem[{\citenamefont{Preskill}()}]{PreskillsNotes}
    \bibinfo{author}{\bibfnamefont{J.}~\bibnamefont{Preskill}}.
    ``Lecture notes on quantum information and computation'',
    \url{www.theory.caltech.edu/people/preskill/ph229/#lecture}.

\bibitem{Lanyon} B.P. Lanyon \etal, Nat. Phys. \textbf{5}, 134 (2009);
T.C. Ralph, K.J. Resch and A. Gilchrist, Phys. Rev. A \textbf{75} 022313 (2007).
    
\bibitem{Schrodinger} E. Schr\"odinger, Proc. Camb. Phil. Soc.
    31, 555 (1935); L.P. Hughston, R. Jozsa and W.K.
    Wootters, Phys. Lett. A \textbf{183}, 14 
    (1993).

\bibitem{BB84} C.H. Bennett and G. Brassard, Proc.
    IEEE International Conf. on Computers, Systems, and
    Signal Processing, IEEE (New York), 175 (1984); R.W.
    Spekkens and T. Rudolph, Quantum Inform. Compu. \textbf{2}, 66
    (2002).

\bibitem{Spek2} T. Rudolph and R.W. Spekkens, Phys. Rev. A \textbf{70},
    052306 (2004).

\bibitem{Pittman}T.B. Pittman, B.C. Jacobs, and J.D. Franson, Phys. Rev. A
\textbf{66}, 052305 (2002).

\bibitem[{\citenamefont{Kim et~al.}(2006)\citenamefont{Kim,
    Fiorentino, and
  Wong}}]{Kim2006}
	\bibinfo{author}{\bibfnamefont{T.}~\bibnamefont{Kim}},
  \bibinfo{author}{\bibfnamefont{M.}~\bibnamefont{Fiorentino}},
  \bibnamefont{and} \bibinfo{author}{\bibfnamefont{F.N.C.}
  \bibnamefont{Wong}}, \bibinfo{journal}{Phys. Rev. A} \textbf{\bibinfo{volume}{73}},
  \bibinfo{eid}{012316} (\bibinfo{year}{2006});
   A. Fedrizzi \etal,
 	 	\bibinfo{journal}{Optics Express} \textbf{\bibinfo{volume}{15}},
  	\bibinfo{pages}{15377} (\bibinfo{year}{2007}).


\bibitem[{\citenamefont{O'Brien
    et~al.}(2003)\citenamefont{O'Brien, Pryde,
  White, Ralph, and Branning}}]{Obrien2003}
	J.L. O'Brien \etal,
	  \bibinfo{journal}{Nature} \textbf{\bibinfo{volume}{426}},
	  \bibinfo{pages}{264} (\bibinfo{year}{2003}).

\bibitem{White2001}A.G. White \etal, Phys. Rev. A.
	\textbf{65}, 012301 (2001).

\bibitem{Jozsa} Definition: 
	$F(\rho,\sigma)=\left[ \mathrm{Tr}\left( \sqrt{\sqrt{\rho}\sigma \sqrt{\rho}}\right)\right]^2$. 
	R. Jozsa, J. Modern Optics \textbf{41}, 2315 (1994).

\bibitem[{\citenamefont{James et~al.}(2001)\citenamefont{James,
    Kwiat, Munro,
 	 and White}}]{James2001}
	D.F.V. James \etal,
	  \bibinfo{journal}{Phys. Rev. A} \textbf{\bibinfo{volume}{64}},
	  \bibinfo{pages}{052312} (\bibinfo{year}{2001}).
	  
\end{thebibliography}
\end{document}